\documentclass[10pt]{article}

\usepackage{graphicx}
\usepackage{amsmath}
\usepackage{amssymb}

\numberwithin{equation}{section}
\newtheorem{thm}{Theorem}[section]

\newcommand{\eqa}{\begin{eqnarray}}
\newcommand{\eeqa}{\end{eqnarray}}
\newcommand{\beq}{\begin{equation}}
\newcommand{\eeq}{\end{equation}}
\newcommand{\nn}{\nonumber}
\newcommand{\pp}{\partial}
\newcommand{\VE}{\varepsilon}

\newcommand{\lm}{\lambda}

\begin{document}
\parskip 6pt
\hoffset -1.8cm

\title{Geometric Flows of Curves, Two-Component  Camassa-Holm Equation and Generalized Heisenberg Ferromagnet Equation}
\author{Aigul Taishiyeva\footnote{Email: ataishiyeva@gmail.com},  \,           Tolkynay Myrzakul\footnote{Email: trmyrzakul@gmail.com}, \, Gulgassyl  Nugmanova\footnote{Email: nugmanovagn@gmail.com},  \, Shynaray Myrzakul\footnote{Email: srmyrzakul@gmail.com}, \, \\ Kuralay  Yesmakhanova\footnote{Email: kryesmakhanova@gmail.com} \, and Ratbay Myrzakulov\footnote{Email: rmyrzakulov@gmail.com}\\
\textsl{Eurasian International Center for Theoretical Physics and} \\ { Department of General \& Theoretical Physics}, \\ Eurasian National University,
Nur-Sultan, 010008, Kazakhstan
}
\date{}
\maketitle

\begin{abstract}In this paper, we study the generalized Heisenberg ferromagnet equation, namely,  the M-CVI equation. This equation is integrable. The integrable motion of the space curves induced by the M-CVI equation is presented.  Using this result,  the Lakshmanan (geometrical) equivalence  between  the M-CVI equation and the two-component Camassa-Holm equation is established.

\vskip 0.5 cm
\noindent {\bf Mathematics Subject Classifications(2000)}. 35Q53, 37K35\newline
{\bf Key words}: Camassa-Holm equation, AKNS hierarchy, reciprocal transformation, Heisenberg ferromagnet equation, spin systems, motion of space curves, differential geometry of curves and surfaces.
\end{abstract}
\section{Introduction}

The celebrated Camassa-Holm equation (CHE) has the form 
\begin{eqnarray}
u_t+\kappa\, u_x-u_{xxt}+3 u u_x=2 u_x u_{xx}+u u_{xxx},
\label{1.1}
\end{eqnarray}
where $u=u(x,t)$ is the fluid velocity in the $x$ direction and  $\kappa=const$ is related to the critical
shallow water wave speed. 
The CHE shares most of the important properties of 
integrable equations like  the $N$-soliton solutions, the
bi-Hamiltonian structure,  the Lax representation (LR) and so on. In the case, when $\kappa=0$, the CHE
(\ref{1.1}) has the so-called peakon solutions. Several important generalizations of the CHE  including integrable cases  but
many other (non-integrable or whose integrability has not been determined) have been discovered \cite{0501028}-\cite{0805.4310}. In particular, the two-component CHE (2-CHE) was constructed.  Our main interest in this paper is to go further with the investigation initiated in our previous papers (see, e.g. \cite{assem2}-\cite{bayan2}).  In this paper, we study the 2-CHE, its relation with the geometry of space curves and the equivalent spin system. 

The paper is organized as follows. In Section 2 we present main fact for the M-CVI equation. Basic information on the 2-CHE we give in  Section 3. The integrable motion of space curves induced by the M-CVI equation and the 2-CHE is studied in Section 4.  In
Section 4,  we consider the gauge equivalence between the M-CVI equation and the 2-CHE.  Finally, in
Section 5 we present a discussion of our achievements and how they impact some recent results found in
the recent literature.

\section{M-CVI equation}
There are several integrable and non-integrable generalized Heisenberg ferromagnet equations (gHFE) (see, e.g. \cite{R13}-\cite{bayan2}). In this paper, we consider one of the gHFE, namely, the M-CVI equation. The M-CVI equation is integrable. It  shares most of the important properties of 
integrable systems like the Lax representation (LR),  the
bi-hamiltonian structure, the $N$-soliton solutions, infinite hierarchy of symmetries and conservation laws and so on. The M-CVI equation also can admits  the so-called peakon solutions. 
 
\subsection{Equation}
Consider the M-CVI equation
\begin{eqnarray}
 [A,A_{xt}+(uA_{x})_{x}]-\frac{1}{\beta^{2}}A_{x}-4\beta\rho\rho_{x}Z=0.\label{2.1}
\end{eqnarray}
Here  $m=\frac{\det{(A_{x}^{2})}}{4\beta^{2}}=u-u_{xx},\quad \rho^{2}=-\frac{tr(A_{x}^{2})+2det(A_{x})}{8\beta^{4}}, \quad u=0.25\beta^{-2}(1-\partial_{x}^{2})^{-1}\det{(A_{x}^{2})}$ are some real functions, $\beta=const$ and 
\begin{eqnarray}
 Z&=&\frac{0.5\beta}{u_{x}+u_{xx}}[A,A_{t}+( u-0.5\beta^{-2})A_{x}], \quad {\bf A}=(A_{1}, A_{2}, A_{3}), \label{11}\\
A&=&\left(\begin{array} {cc} A_{3}& A^{-} \\ 
A^{+}&-A_{3}\end{array}\right), \quad A^{\pm}=A_{1}\pm iA_{2}, \quad A^{2}=I, \quad {\bf A}^{2}=1.\label{12}
\end{eqnarray}
\subsection{Lax representation}
The  LR of the M-CVI equation reads as
\begin{eqnarray}
\Psi_{x}&=&U_{1}\Psi,\label{2.4}\\
 \Psi_{s}&=&V_{1}\Psi. \label{2.5}
\end{eqnarray}
Here
\begin{eqnarray}
U_{1}&=&\left(\frac{\lambda}{4\beta}-\frac{1}{4}\right)[A,A_{x}]+(\lambda^{3}-\beta^{2}\lambda)\rho^{2}Z,\label{2.6}\\
V_{1}&=&\left(\frac{1}{4\beta^{2}}-\frac{1}{4\lambda^{2}}\right)A+\frac{u}{4}\left(\frac{\beta}{\lambda}-\frac{\lambda}{\beta}\right)[A,A_{x}]+\left(\frac{\beta}{4\lambda}-\frac{1}{4}\right)[A,A_{t}]+v\rho^{2}Z \label{2.7}
\end{eqnarray}
where $v=\lambda(0.5+\beta^{2}u)-\lambda^{3}u-0.5\beta^{2}\lambda^{-1}$. The compatibility condition $U_{1t}-V_{1x}+[U_{1},V_{1}]=0$ is equivalent to the M-CVI equation (\ref{2.1}).

\subsection{Reductions}
One of the reductions of the M-CVI equation is the so-called M-CIV equation. Let $\rho=0$. Then the M-CVI equation takes the form
\begin{eqnarray}
 [A,A_{xt}+(uA_{x})_{x}]-\frac{1}{\beta^{2}}A_{x}=0.\label{10}
\end{eqnarray}
It is nothing but the M-CIV equation (see, e.g. \cite{assem2}-\cite{bayan2}). The LR of the M-CIV equation follows from the LR of the M-CVI equation  (\ref{2.4})-(\ref{2.5}) as $\rho=0$. So that the LR of the M-CIV equation is given by \cite{assem2}
\begin{eqnarray}
\Psi_{x}&=&U_{3}\Psi,\label{2.9}\\
 \Psi_{s}&=&V_{3}\Psi, \label{2.10}
\end{eqnarray}
where
\begin{eqnarray}
U_{3}&=&\left(\frac{\lambda}{4\beta}-\frac{1}{4}\right)[A,A_{x}],\label{2.11}\\
V_{3}&=&\left(\frac{1}{4\beta^{2}}-\frac{1}{4\lambda^{2}}\right)A+\frac{u}{4}\left(\frac{\beta}{\lambda}-\frac{\lambda}{\beta}\right)[A,A_{x}]+\left(\frac{\beta}{4\lambda}-\frac{1}{4}\right)[A,A_{t}]. \label{2.12}
\end{eqnarray}

\section{2-CHE}

\subsection{Equation}

The two-component CHE (2-CHE) is given by \cite{0501028}
\begin{eqnarray}
m_t+um_x+2mu_x-\rho\rho_x&=&0,\label{3.1}\\
\rho_t+(\rho u)_x&=&0,\label{3.2}
\end{eqnarray}
where $m=u-u_{xx}+0.5\kappa$. If  $\rho=0$,  the 2-CHE reduces to the CHE (\ref{1.1}). 
\subsection{Lax representation}
The LR of the 2-CHE is given by (see, e.g.  \cite{0501028})
\begin{eqnarray}
\phi_{xx}&=&(\frac{1}{4}-m\varsigma+\rho^2\varsigma^{2})\phi,\label{3.3}\\
\phi_t&=&-(\frac{1}{2\varsigma}+u)\phi_x+\frac{u_x}{2}\phi,\label{3.4}
\end{eqnarray}
where $\varsigma$ is a spectral parameter and $m=u-u_{xx}+0.5\kappa \quad (\kappa=const$).

\subsection{Reciprocal transformation}

From the equation (\ref{3.2}) follows that  the  $1$-form
\beq
\omega=\rho\,dx-\rho\,u\,dt
\eeq
is closed. This means that we can  define a reciprocal transformation $(x,t)\mapsto (y,s)$ by the relation \cite{0501028}
\begin{equation}
dy=\rho\,dx-\rho\,u\,dt,\quad ds=dt.
\end{equation}
So  we obtain
\begin{equation}\label{zh9}
\frac{\pp}{\pp x}=\rho\,\frac{\pp}{\pp y},\quad \frac{\pp}{\pp t}=\frac{\pp}{\pp s}-\rho\,u\,\frac{\pp}{\pp y}.
\end{equation}
Then the spectral problem (\ref{3.3})-(\ref{3.4}) takes the form \cite{0501028}
\begin{eqnarray}
\varphi_{yy}&=&\left(\lm^2-P\,\lm-Q\right)\varphi,\label{3.8}\\
\varphi_s&=&-\frac{\rho}{2\lm}\varphi_y+\frac{\rho_y}{4\lm}\varphi,\label{3.9}
\end{eqnarray}
where
\beq
\varphi=\sqrt{\rho}\,\phi, \quad 
P=\frac{m}{\rho^2},\quad
Q=-\frac1{4\rho^2}-\frac{\rho_{yy}}{2\rho}+\frac{\rho_y^2}{4\rho^2}.\label{tr-2}
\eeq
The compatibility condition of the equations (\ref{3.8})-(\ref{3.9}) gives
\begin{eqnarray}
&&P_s=\rho_y, \label{ed-1}, \label{3.11} \\
&&Q_s+\frac12\rho\,P_y+P\,\rho_y=0, \label{ed-2} \\
&&\frac12\rho\,Q_y+Q \rho_y+\frac14\rho_{yyy}=0. \label{ed-3}
\end{eqnarray}
Hence we get the equation \cite{0501028}
\begin{equation}
\rho^2\,Q+\frac12\rho\,\rho_{yy}-\frac14\rho_y^2=C=-\frac14.
\end{equation}
From the equation (\ref{3.11}) follows 
\begin{equation} \label{tr-3}
P=\frac{\pp f(y,s)}{\pp y},\ \rho=\frac{\pp f(y,s)}{\pp s},
\end{equation}
where $f(y,s)$ is some function. This function satisfies the equation \cite{0501028}
\begin{equation}\label{ef}
\frac{f_{ss}}{2f_s^3}+f_yf_{ys}-\frac{f_{ss}f_{ys}^2}{2f_s^3}+\frac{f_{ys}f_{yss}}{2f_s^2}
+\frac12f_sf_{yy}+\frac{f_{ss}f_{yys}}{2f_s^2}-\frac{f_{yyss}}{2f_s}=0.
\end{equation}
Finally we come to the following theorems \cite{0501028}
\begin{thm}\label{fuv}
Let $f$ be a solution of the equation (\ref{ef}), and
\begin{equation}
u=f_yf_s^2+\frac{f_{ss}f_{ys}}{f_s}-f_{yss},\quad \rho=f_s. \label{zh6}
\end{equation}
If $x(y,s)$ is a solution of the following system of ODEs:
\begin{equation}
\frac{\pp x}{\pp y}=\frac1{\rho},\quad  \frac{\pp x}{\pp s}=u, \label{zh7}
\end{equation}
then $(u(y,t),\rho(y,t),x(y,t))$ is a parametric solution of the 2-CHE (\ref{3.1})-(\ref{3.2}).
\end{thm}
\begin{thm}\label{fx}
Let $f(y,s)$ be a solution of the equation (\ref{ef}). Define the functions $x=x(y,s), u=u(y,s), \rho=\rho(y,s)$ by
\begin{equation}\label{zh8}
x=f(s,y),\quad u=\frac{\pp x}{\pp s},\quad \frac1{\rho}=\frac{\pp x}{\pp y}.
\end{equation}
Then $(u(y,t),\rho(y,t),x(y,t))$ is a parametric solution of the 2-CHE (\ref{3.1})-(\ref{3.2}).
\end{thm}

\subsection{Relations to the first negative flow of the AKNS hierarchy}
Let us briefly present the well-known result about relation between the 2-CHE and the AKNS spectral problem following the paper \cite{0501028}. 
The AKNS spectral problem reads as 
\begin{equation}
\left(\begin{array}{c} \phi_1 \\ \phi_2 \end{array}\right)_y=
\left(\begin{array}{rr} \lm & -q \\ r & -\lm \end{array}\right)
\left(\begin{array}{c} \phi_1 \\ \phi_2 \end{array}\right).\label{3.20}
\eeq
The first negative flow of the AKNS problem is given by
\beq
\left(\begin{array}{c} \phi_1 \\ \phi_2 \end{array}\right)_s=
\frac1{4\lm}\left(\begin{array}{rr} a & b \\ c & -a \end{array}\right)
\left(\begin{array}{c} \phi_1 \\ \phi_2 \end{array}\right). \label{3.21}
\end{equation}
The compatibility condition of the equations (\ref{3.20})-(\ref{3.21}) gives
\begin{eqnarray}
&&q_s=\frac12\, b, \\
&&r_s=\frac12\, c, \label{ak-1}\\
&&b_y=2\,a\,q,\\
&&c_y=2\,a\,r, \label{ak-2}\\
&&a_y+b\,r+c\,q=0.\label{ak-3}
\end{eqnarray}
Hence we get the condition \cite{0501028}
\begin{equation}\label{kk}
a^2+b\,c=\VE^2,
\end{equation}
where $\VE=const$. We have the following theorems \cite{0501028}:
\begin{thm}\label{qrf}
Let $(a,b,c,q,r)$ be a solution of the equations (\ref{ak-1})--(\ref{ak-3}) with $\VE^2=1$,
then any function $f(y,s)$ satisfying
\begin{equation}\label{ff-1}
2a=b\,e^{-f}-c\,e^f
\end{equation}
gives a primary solution of the 2-CH system.
\end{thm}
\begin{thm}\label{fqr}
If $f$ is a primary solution of the 2-CHE system (\ref{3.1})-(\ref{3.2}),
then we can construct a solution of the first negative flow of the AKNS hierarchy by the following formulae
\begin{equation}
q=\frac{e^f}2\left(f_y+\frac{\VE-f_{ys}}{f_s}\right),\ r=\frac{e^{-f}}2\left(f_y-\frac{\VE-f_{ys}}{f_s}\right),\
b=2\,q_s,\ c=2\,r_s,\ a=\frac{b\,e^{-f}-c\,e^f}2.
\end{equation}
where $\VE=1$ or $\VE=-1$.
\end{thm}

\subsection{Bi-Hamiltonian structure}
Let us again briefly present some basic facts about the bi-Hamiltonian structure of the 2-CHE following the paper \cite{0501028}. Note  
that both bi-Hamiltonian structures of the CHE  and  the KdV hierarchies
are deformations of the following bi-Hamiltonian structure
of hydrodynamic type \cite{0501028}
\begin{eqnarray}
&&\{u(x),u(y)\}_1=\delta'(x-y),\nn\\
&&\{u(x),u(y)\}_2=u(x)\delta'(x-y)+\frac12\,u(x)'\delta(x-y).\label{z8}
\end{eqnarray}
This means  that the dispersionless limits of the CHE and 
KdV hierarchies have the same forms. One of the main features
of the integrable hierarchies that correspond to bi-Hamiltonian structures with constant central invariants
is the existence of $\tau$-functions. It is  well known that integrable hierarchies of nonlinear  partial differential equations 
with one spatial variable possess bi-Hamiltonian structures that are deformations of
bi-Hamiltonian structure of hydrodynamic type with constant central invariants, and the existence
of $\tau$-functions plays an important role in the study of these integrable systems. The CHE hierarchy is an exceptional example of integrable systems which does not possess $\tau$-functions.

\section{Motion of space curves induced by the M-CIV equation. Lakshmanan (geometrical) equivalence}

In this section, we would like to find the integrable motion of the space curves induced by the M-CVI equation. To do that, let us consider a smooth space curve in $R^{3}$ given by 
\begin{eqnarray}
{\bf \gamma} (x,t): [0,X] \times [0, T] \rightarrow R^{3},\label{17}
\end{eqnarray} 
where $x$ is the arc length of the curve at each time $t$.  Then the  following three vectors 
\begin{eqnarray}
{\bf e}_{1}={\bf \gamma}_{x}, \quad {\bf e}_{2}=\frac{{\bf \gamma}_{xx}}{|{\bf \gamma}_{xx}|}, \quad {\bf e}_{3}={\bf e}_{1}\wedge {\bf e}_{2}, \label{48}
\end{eqnarray} 
are the  unit tangent vector, the  principal normal vector and the binormal vector of the curve, respectively.
 The   corresponding Frenet-Serret equation is given by
 \begin{eqnarray}
\left ( \begin{array}{ccc}
{\bf  e}_{1} \\
{\bf  e}_{2} \\
{\bf  e}_{3}
\end{array} \right)_{x} = C
\left ( \begin{array}{ccc}
{\bf  e}_{1} \\
{\bf  e}_{2} \\
{\bf  e}_{3}
\end{array} \right)=
\left ( \begin{array}{ccc}
0   & \kappa_{1}     & \kappa_{2} \\
-\kappa_{1}& 0     & \tau  \\
-\kappa_{2}   & -\tau & 0
\end{array} \right)\left ( \begin{array}{ccc}
{\bf  e}_{1} \\
{\bf  e}_{2} \\
{\bf  e}_{3}
\end{array} \right), \label{49} 
\end{eqnarray}
where $\tau$,  $\kappa_{1}$ and $\kappa_{2}$ are   torsion,  geodesic curvature and  normal curvature of the curve, respectively. 
Let the   deformation of the curves  are given by 
\begin{eqnarray}
\left ( \begin{array}{ccc}
{\bf  e}_{1} \\
{\bf  e}_{2} \\
{\bf  e}_{3}
\end{array} \right)_{x} = C
\left ( \begin{array}{ccc}
{\bf  e}_{1} \\
{\bf  e}_{2} \\
{\bf  e}_{3}
\end{array} \right),\quad
\left ( \begin{array}{ccc}
{\bf  e}_{1} \\
{\bf  e}_{2} \\
{\bf  e}_{3}
\end{array} \right)_{t} = G
\left ( \begin{array}{ccc}
{\bf  e}_{1} \\
{\bf  e}_{2} \\
{\bf  e}_{3}
\end{array} \right). \label{4.4} 
\end{eqnarray}
Here
\begin{eqnarray}
C &=&-\tau L_{1}+\kappa_{2}L_{2}-\kappa_{1}L_{3}=
\left ( \begin{array}{ccc}
0   & \kappa_{1}     & \kappa_{2}  \\
-\kappa_{1}  & 0     & \tau  \\
-\kappa_{2}    & -\tau & 0
\end{array} \right) ,\\
G &=&-\omega_{1}L_{1}+\omega_{2}L_{2}-\omega_{3}L_{3}=
\left ( \begin{array}{ccc}
0       & \omega_{3}  & \omega_{2} \\
-\omega_{3} & 0      & \omega_{1} \\
-\omega_{2}  & -\omega_{1} & 0
\end{array} \right),\label{51} 
\end{eqnarray}
where
\begin{eqnarray}
 L_{1} = \begin{bmatrix}0&0&0\\0&0&-1\\0&1&0\end{bmatrix} , \quad
 L_{2} = \begin{bmatrix}0&0&1\\0&0&0\\-1&0&0\end{bmatrix} , \quad
 L_{3} = \begin{bmatrix}0&-1&0\\1&0&0\\0&0&0\end{bmatrix}
\end{eqnarray}
are basis elements of $so(3)$ algebra.
The compatibility condition of the equations (\ref{4.4}) has the form
\begin{eqnarray}
C_t - G_x + [C, G] = 0\label{4.8} 
\end{eqnarray}
or in components   
 \begin{eqnarray}
\kappa_{1t}- \omega_{3x} -\kappa_{2}\omega_{1}+ \tau \omega_2&=&0, \label{4.9} \\ 
\kappa_{2t}- \omega_{2x} +\kappa_{1}\omega_{1}- \tau \omega_3&=&0, \label{4.10} \\
\tau_{t}  -    \omega_{1x} - \kappa_{1}\omega_2+\kappa_{2}\omega_{3}&=&0.  \label{4.11} \end{eqnarray}
As usual, we consider  the following identification ${\bf A}\equiv {\bf e}_{1}$. We have     
\begin{eqnarray}
\kappa_{1}=i, \quad \kappa_{2}=\lambda(m-1)+\lambda^{3}\rho^{2}, \quad \tau=-i[\lambda(m+1)+\lambda^{3}\rho^{2}], \label{57} 
\end{eqnarray}
where $\kappa_{2}+i\tau=2m\lambda+2\rho^{2}\lambda^{3}$ and $\kappa_{2}-i\tau=-2\lambda$,  $\lambda=const$.  Finally we get the following expressions for the functions $\omega_{j}$ 
\begin{eqnarray}
\omega_{1} & = &i[(u\lambda-0.5\lambda^{-1})(m+1)-0.5\lambda^{-1}(u_{x}+u_{xx})-0.5\lambda\rho^{2}+\lambda^{3}u\rho^{2}],\label{58}\\ 
\omega_{2}&=& [(0.5\lambda^{-1}-\lambda u)(m-1)+0.5\lambda^{-1}(u_{x}+u_{xx})+0.5\lambda\rho^{2}-\lambda^{3}u\rho^{2}], \label{59} \\
\omega_{3} & = &i[0.5\lambda^{-2}-u-u_{x}].      \label{60}
\end{eqnarray}
Substituting these expressions to Eqs.(\ref{4.9})-(\ref{4.11}) we get the following equations for $m, \rho$: 
\begin{eqnarray}
m_{t}+2u_{x}m+um_{x}-\rho\rho_{x}&=&0, \label{33} \\
\rho_{t}+(u\rho)_{x}&=&0, \label{9}\\
m-u+u_{xx}-0.5\kappa&=&0, \label{62} \label{34}
\end{eqnarray}
which is the 2-CHE. 
So, we have  proved  the  Lakshmanan (geometrical) equivalence between the M-CVI  equation  (\ref{2.1}) and the 2-CHE (\ref{3.1})-(\ref{3.2}).

\section{Gauge equivalence}
In the previous section, we have shown that  the M-CIV equation (\ref{2.1}) and the 2-CHE (\ref{3.1})-(\ref{3.2}) are  the geometrical equivalent each to other. As it was established in \cite{aigul1},  between these equations takes place  also the gauge equivalence. In fact, consider the gauge transformation
$\Phi=g\Psi$, where $g=\Phi|_{\lambda=\beta}$. Then we have
\begin{eqnarray}
U_{1}=g^{-1}U_{2}g-g^{-1}g_{x}, \quad V_{1}=g^{-1}V_{2}g-g^{-1}g_{t}.\label{5.1} 
\end{eqnarray}
As result, we get the following LR for the 2-CHE
\begin{eqnarray}
\Phi_{x}&=&U_{2}\Phi, \label{5.2}\\
\Phi_{t}&=&V_{2}\Phi,\label{5.3} 
\end{eqnarray}
where 
\begin{eqnarray}
U_{2} &=&\left ( \begin{array}{cc}
-0.5   & \lambda  \\
m\lambda+\rho^{2}\lambda^{3}  & 0.5
\end{array} \right), \label{5.4}\\
V_{2} &=&\left ( \begin{array}{cc}
0.5(u+u_{x})-0.25\lambda^{-2}       & 0.5\lambda^{-1}-u\lambda \\
0.5(m+u_{x}+u_{xx})\lambda^{-1}-um+0.5\rho^{2}\lambda-u\rho^{2}\lambda^{3} & 0.25\lambda^{-2}-0.5(u+u_{x})
\end{array} \right).\label{5.5} 
\end{eqnarray}
The compatibility condition 
\begin{eqnarray}
U_{2t}-V_{2x}+[U_{2},V_{2}]=0\label{5.6} 
\end{eqnarray}
gives the 2-CHE.

\section{Conclusion}
In this paper, we have considered the M-CVI equation and the 2-CHE. First, we have presented some well-known main facts  on these equations. In particular, we briefly present 
the  reciprocal transformation between the 2-CHE  and the first negative
flow of the AKNS hierarchy which includes in particular the well
known sine-Grodon and the sinh-Gordon equations. This transformation gives the  correspondence between solutions of the first negative
flow of the AKNS hierarchy and the 2-CHE.
Then we have studied the motion of the space curves induced by these equations. Using this result, we have proved that the M-CVI equation and the 2-CHE is the Lakshmanan (geometrical) equivalent each to other. Last but not least, we would like to note and believe that the "spinalization" of integrable systems gives some new informations on their  nature. 

\vskip 0.2truecm \noindent{\bf Acknowledgments.} This work was supported  by  the Ministry of Edication  and Science of Kazakhstan under
grants 0118РК00935 and 0118РК00693.

\end{document}